\documentclass[sigconf]{acmart}
\usepackage{arydshln}
\usepackage{caption}
\usepackage{multirow}
\usepackage{siunitx}
\usepackage{nameref}

\makeatletter
\renewcommand{\email}[2][]{%
  \g@addto@macro\addresses{}
}
\makeatother

\AtBeginDocument{%
  }

\setcopyright{acmlicensed}
\copyrightyear{2026}
\acmYear{2026}
\setcopyright{cc}
\setcctype{by}
\acmConference[ICSE '26]{2026 IEEE/ACM 48th International Conference on Software Engineering}{April 12--18, 2026}{Rio de Janeiro, Brazil}
\acmBooktitle{2026 IEEE/ACM 48th International Conference on Software Engineering (ICSE '26), April 12--18, 2026, Rio de Janeiro, Brazil}
\acmPrice{}
\acmDOI{10.1145/3744916.3773124}
\acmISBN{979-8-4007-2025-3/26/04}

\begin{document}

\title[Staying or Leaving?]{\looseness=-1 Staying or Leaving? How Job Satisfaction, Embeddedness and Antecedents Predict Turnover Intentions of Software Professionals}

%


\author{Miikka Kuutila}
\email{miikka.kuutila@helsinki.fi}
\affiliation{%
  \institution{Dalhousie University}
  \city{Halifax}
  \country{Canada}
}
\affiliation{%
  \institution{University of Helsinki}
  \country{Finland}
  \texttt{miikka.kuutila@\,helsinki.fi}
}

\author{Paul Ralph}
\email{paulralph@dal.ca}
\affiliation{%
  \institution{Dalhousie University}
  \city{Halifax}
  \country{Canada}
  \texttt{paulralph@\,dal.ca}
}

\author{Huilian Sophie Qiu}
\email{sophie.qiu@kellogg.northwestern.edu}
\affiliation{%
  \institution{Northwestern University}
  \city{Evanston}
  \country{USA}
  \texttt{huilian\_qiu@}\\
  \texttt{\,alumni.brown.edu}\\
}

\author{\mbox{Ronnie de Souza Santos}}
\email{ronnie.desouzasantos@ucalgary.ca}
\affiliation{%
  \institution{University of Calgary}
  \city{Calgary}
  \country{Canada}
  \texttt{ronnie.desouzasantos}
  \texttt{@ucalgary.ca}
}

\author{Morakot Choetkiertikul}
\email{morakot.cho@mahidol.ac.th}
\affiliation{%
  \institution{Mahidol University}
  \city{Nakhon Pathom}
  \country{Thailand}
  \texttt{morakot.cho@\,mahidol.ac.th}
}

\author{Amin Milani Fard}
\email{amilanif@nyit.edu}
\affiliation{%
  \institution{New York Institute of Technology}
  \city{Vancouver}
  \country{Canada}
  \texttt{amilanif@\,nyit.edu}
}

\author{Rana Alkadhi}
\email{ralkadi@KSU.EDU.SA}
\affiliation{%
  \institution{King Saud University}
  \city{Riyadh}
  \country{Saudi Arabia}
  \texttt{ralkadi@\,ksu.edu.sa}
}

\author{Xavier Devroey}
\email{xavier.devroey@unamur.be}
\orcid{0000-0002-0831-7606}
\affiliation{%
  \institution{NADI, University of Namur}
  \city{Namur}
  \country{Belgium}
  \texttt{xavier.devroey@\,unamur.be}
}

\author{Gregorio Robles}
\email{grex@gsyc.urjc.es}
\affiliation{%
  \institution{U. Rey Juan Carlos}
  \city{Madrid}
  \country{Spain}
  \texttt{grex@\,gsyc.urjc.es}
}

\author{Hideaki Hata}
\email{hata@shinshu-u.ac.jp}
\affiliation{%
  \institution{Shinshu University}
  \city{Nagano}
  \country{Japan}
  \texttt{hata@\,shinshu-u.ac.jp}
}

\author{Sebastian Baltes}
\email{research@sbaltes.com}
\affiliation{%
  \institution{Heidelberg University}
  \city{Heidelberg}
  \country{Germany}
  \texttt{sebastian.baltes@}
  \texttt{\,uni-heidelberg.de}
}

\author{Hera Arif}
\email{hera.arif@dal.ca}
\affiliation{%
  \institution{Dalhousie University}
  \city{Halifax}
  \country{Canada}
  \texttt{hera.arif@\,dal.ca}
}

\author{Vladimir Kovalenko}
\email{vladimir.kovalenko@jetbrains.com}
\affiliation{%
  \institution{JetBrains, Amsterdam}
  \country{The Netherlands}
  \texttt{vladimir.kovalenko@}
  \texttt{jetbrains.com}
}

\author{Shalini Chakraborty}
\email{shalini.chakraborty@uni-bayreuth.de}
\affiliation{%
  \institution{University of Bayreuth}
  \city{Bayreuth}
  \country{Germany}
  \texttt{shalini.chakraborty}
  \texttt{@uni-bayreuth.de}
}

\author{Eray Tuzun}
\email{eraytuzun@cs.bilkent.edu.tr}
\affiliation{%
  \institution{Bilkent University}
  \city{Ankara}
  \country{Turkey}
  \texttt{eraytuzun@\,cs.bilkent.edu.tr}
}

\author{Gianisa Adisaputri}
\email{gianisa@dal.ca}
\affiliation{%
  \institution{Dalhousie University}
  \city{Halifax}
  \country{Canada}
  \texttt{gianisa@\,dal.ca}
}

\renewcommand{\shortauthors}{Kuutila et al.}




\begin{abstract}
\textbf{Context:} Voluntary turnover is common in the software industry, increasing recruitment and onboarding costs and the risk of losing organizational and tacit knowledge.
\textbf{Objective:} This study investigates how job satisfaction, work-life balance, job embeddedness, and their antecedents, including job quality, personality traits, attitudes toward technical and sociotechnical infrastructure, and perceptions of organizational justice, relate to software professionals' turnover intentions.
\textbf{Method:} We conducted a geographically diverse cross-sectional survey of software professionals (N = 224) and analyzed the data using partial least squares structural equation modeling (PLS-SEM). Our model includes both reflective and formative constructs and tests 15 hypotheses grounded in occupational psychology and software engineering literature. Additionally, indirect effects were explored.
\textbf{Results:} Job satisfaction and embeddedness were significantly negatively associated with software professionals' turnover intentions, while work-life balance showed no direct effect. The strongest antecedents for job satisfaction were work-life balance and job quality, while organizational justice was the strongest predictor of job embeddedness. Exploratory analysis suggests that job satisfaction and job embeddedness mediate the effect of job quality, organizational justice, and work-life balance on turnover intentions.
\textbf{Discussion:} The resulting PLS-SEM model has considerably higher explanatory power for key outcome variables than previous work conducted in the software development context, highlighting the importance of both psychological (e.g., job satisfaction, job embeddedness) and organizational (e.g., organizational justice, job quality) factors in understanding turnover intentions of software professionals. Our results imply that improving job satisfaction and job embeddedness is the key to retaining software professionals. In turn, enhancing job quality, supporting work-life balance, and ensuring high organizational justice can improve job satisfaction and embeddedness, indirectly reducing turnover intentions. Our results broadly suggest that retention strategies are best focused on fairness, meaningful work, and work-life balance.
\end{abstract}

\maketitle

\section{Introduction}
Voluntary employee turnover (i.e., employees choosing to leave their jobs) damages organizational effectiveness because knowledge and skills are lost~\cite{abboud2020}. This knowledge loss is particularly harmful in the software industry because expert knowledge can be tacit~\cite{ryan2013acquiring}, making it harder to document and share. Moreover, both recruiting~\cite{huang2009synthesizing} and onboarding~\cite{britto2018onboarding} software professionals is expensive and difficult because of the specificity of their skills and the novelty and complexity of the workflows to which they must adjust to.

Much of the turnover literature focuses on process models~\cite{steel2009turnover} and the connection between voluntary turnover and job satisfaction~\cite{steel2009turnover, da2025decade}. While these models offer valuable insights into general turnover dynamics, they may not adequately capture the unique conditions and pressures of the software industry. Indeed, seminal review articles on turnover call for industry-specific studies and models~\cite{rubenstein2018surveying}, which motivates the following research question.

\smallskip
{\narrower \noindent \textit{\textbf{Research Question:} What are the key psychological, organizational, and job-related factors that shape turnover intentions among software professionals?}\par}
\smallskip


To answer this research question, we first familiarized ourselves with central concepts and prior literature regarding voluntary turnover, which we outline in Section~\ref{sec:background}. Based on the review of concepts related to turnover identified from prior literature, we developed 15 hypotheses to be tested as shown in Section~\ref{sec:model}. We tested these hypotheses with data from a survey instrument consisting mostly of previously validated, widely used psychometric scales. The methodology we used to analyze 224 responses to our survey instrument is explained in Section~\ref{sec:methodology}, and the results are presented in Section~\ref{sec:results}. Our results and their implications are discussed in relation to prior work from both software engineering and IT domains in Section~\ref{sec:discussion}. Lastly, we conclude the article with main takeaways in Section~\ref{sec:conclusions}.
\section{Related Work}\label{sec:background}
This section reviews (1) seminal theories in social psychology that explain human behavior; (2) key concepts from the turnover literature; (3) relevant studies from software engineering and the broader field of information technology.

\subsection{Human behavior}
When explaining human behavior, such as changing jobs, two seminal theories, the Theory of Reasoned Action and the Theory of Planned Behavior, are frequently used. These theories posit that behavior is mostly driven by intentions; that is, most people, most of the time, do what they intend unless something gets in the way. Consequently, the intention to change jobs is frequently measured in the turnover literature as an important predictor of actual turnover. Turnover intention is viewed as ``the willingness of the employee to leave her organization within a given period of time''~\cite{lazzari2022predicting}. 

Meanwhile, intentions result in part from individuals' attitudes and perceived subjective norms~\cite{fishbein1977belief, montano2015theory}. Attitudes are ``overall evaluations of objects,'' which consist of ``affective information (e.g., feelings towards an object), cognitive information (e.g., beliefs associated with an object) and behavioral information (e.g., past experiences with an object)''~\cite{haddock2004contemporary}. However, attitudes are more stable over time than core affect~\cite{haddock2004contemporary}, understood as ``a neurophysiological state that is consciously accessible as a simple, nonreflective feeling that is an integral blend of hedonic (pleasure–displeasure) and arousal (activation–deactivation) values''~\cite{russell2003core}. A subjective norm is ``a perception that an individual has regarding whether people important to that individual believe that they should or should not perform a particular behavior''~\cite{apa_dictionary}.


\subsection{Antecedents of voluntary turnover}

Job satisfaction is widely studied in the turnover literature and is one of its main predictors~\cite{hom2017one}. In general, job satisfaction refers to the judgments individuals make about their jobs~\cite{locke1969job, Fernández-Macias2014}. Job satisfaction is inversely related to turnover intentions, that is, satisfied employees are less likely to leave~\cite{hom2017one, lambert2001impact}. Job satisfaction is also associated with better performance, organizational citizenship, and lower absenteeism~\cite{judge2001job}.



Job embeddedness, introduced by Mitchell et al.~\cite{mitchell2001people}, is a construct that encompasses individuals' attachment to their work organization and community, and has been linked to turnover intentions~\cite{jiang2012and}. It is a combination of on-the-job and off-the-job factors that are not about the work itself. Job embeddedness is divided into six dimensions: links, fit, and sacrifice, each for both the work organization and the community. Links refer to connections to other people, fit refers to the perceived alignment with the values of the work organization and the community, and lastly, sacrifice refers to perceived loss resulting from leaving their work organization and their community.

Work-life balance has two key dimensions: role engagement in work and non-work life, and minimal conflict between work and non-work roles~\cite{sirgy2018work}. Prior work on work-life balance supports higher job satisfaction and lower turnover intentions, as employees feel more able to manage responsibilities without sacrificing personal well-being~\cite{sirgy2018work, haar2014outcomes}. 

The Job Characteristics Model generalizes jobs as having five dimensions: autonomy, task identity, skill variety, task significance, and feedback, and is concerned with how employees perceive these factors~\cite{hackman1975development}. Autonomy refers to the freedom and discretion an employee has in scheduling and determining work procedures, while task identity refers to the extent to which an employee completes a whole piece of work from start to finish. Skill variety describes the range of skills a job requires, while task significance refers to the impact the job has on others, both within and beyond the organization. Lastly, feedback refers to the extent to which the work itself informs an individual about how well they are performing. Research indicates that these job characteristics influence both job satisfaction and turnover intentions, highlighting their relevance to understanding voluntary turnover in complex work environments~\cite{nguyen2017organizational, morris2010job}. 

One antecedent of job embeddedness is organizational justice~\cite{nguyen2017organizational}. Organizational justice refers broadly to the perceived fairness in the workplace. Organizational justice has been divided into procedural, interpersonal, and distributive justice~\cite{colquitt2001justice}. Procedural justice refers to perceived fairness in processes and procedures with which decisions are made in a work organization. Interpersonal justice refers to perceived fairness in communications with supervisors, and it concerns the degree to which people feel they are treated with respect and dignity. Distributive justice refers to perceived fairness in allocating resources and distributing rewards, thus, for example, it deals with promotions and pay raises.

The Five-Factor Model of Personality has also been linked to job satisfaction and other turnover antecedents. The meta-analysis by Judge, Heller \& Mount~\cite{judge2002five} examines correlations between Five-Factor Model traits and job satisfaction. Their meta-analysis shows correlations of .29 for neuroticism, .25 for extraversion, .02 for openness to experience, .17 for agreeableness, and .26 for conscientiousness. Five-Factor Model of Personality has also been studied in relation to another antecedent of turnover intentions: job embeddedness. Based on the study by Peltokorpi and Sekiguchi~\cite{peltokorpi2023interaction}, job embeddedness correlates .15 with extraversion, .17 with agreeableness, and .14 with openness to experience.


\subsection{Voluntary turnover research in software engineering and information technology}

A recent review on turnover research in IT by da Silva and Sampaio~\cite{da2025decade} shows that the focus of prior studies in explaining turnover in IT research, that is, not strictly software engineering samples, has been on job satisfaction and organizational commitment. However, many concepts that we reviewed in the previous section are missing from quantitative studies reviewed by da Silva and Sampaio~\cite{da2025decade}. 

In strictly software engineering research, staff turnover has been investigated in the software engineering literature for at least three decades~\cite{abdel1989study}. For example, Abdel-Hamid introduced human resource management into his system dynamics model to investigate the effects of turnover on cost and schedule of software projects, noting that turnover, acquisition and assimilation rates can increase project costs and duration by 40\% to 60\%. Similarly, Foucault et al.~\cite{foucault2015impact} observe decreased software quality with increased turnover in open source software development. 

Bass, Beecham, and Nolte~\cite{bass2018employee} conducted two case studies comparing turnover in in-house and offshore outsourced software development. They identified several factors affecting turnover, including employment policies, workplace innovation, product quality, alignment of offshore and onshore work hours, and excessive working hours with negative health effects. Work–life balance issues were particularly pronounced in the outsourced case.

Sharma and Stol~\cite{sharma2020exploring} investigated the the effect of onboarding success on organizational fit and turnover intentions. The PLS-SEM model fitted with a sample of 102 responses from software professionals indicated that providing support, such as availability of help for new hires played a major role in onboarding success while training and orientation were less important. Additionally, they found that job satisfaction mediated the relationship between onboarding success and turnover intention, but not with workplace quality.

Similarly, Trinkenreich, Santos, and Stol~\cite{trinkenreich2024predicting} used PLS-SEM to model intent to stay at a job by software professionals. Among other results, their model was informed by the Job Demands and Resources model achieved $R^2 = 0.335$ for explaining variance in turnover intentions. Another recent study showed that calculative commitment (i.e., committing based on calculation of benefits and losses) was more important for software engineers than affective commitment (i.e, committing based on liking the organization)~\cite{lewicka2024really}.
\section{Hypotheses}\label{sec:model}

Based on the literature reviewed above, we specify hypotheses and explain our modeling decisions in this section.

\subsection{Measurement models and their validation}\label{ssec:mmv}

Briefly, philosophical realism holds that rather than directly perceiving phenomena, scientists mostly make inferences about phenomena by analyzing data collected using instruments. Since these instruments depend on theories and values, scientists must question the validity of their observations, typically using a measurement model in which properties of target phenomena are modeled as \textit{latent constructs} (e.g., job satisfaction), which are inferred by aggregating \textit{observable variables} (e.g., responses to questions)~\cite{ralph2024teaching}. These latent constructs may be conceptualized in two ways~\cite{becker2012hierarchical, ralph2024teaching}. A \textit{reflective} latent construct is viewed as causing the indicators and estimated by their shared variance~\cite{freeze2007assessment}. In contrast, \textit{formative} latent constructs are composed of or caused by their indicators~\cite{ralph2024teaching, becker2012hierarchical}, which may or may not be correlated. Our proposed theory includes both formative and reflective constructs, as explained below.


\subsection{Hypotheses and modeling decisions}
Since turnover \textit{behavior} can only be measured retrospectively or longitudinally, and turnover intention has about a 0.5 correlation with turnover behavior~\cite{steel1984review}, many studies focus on intention (e.g.~\cite{sharma2020exploring, trinkenreich2024predicting}). Likewise, we adopt turnover intention as our main endogenous variable and model it reflectively (since intentions are a unidimensional psychological phenomenon).

Since job satisfaction is among the most-studied predictors of voluntary turnover~\cite{hom2017one}, we hypothesize that it reduces turnover intentions (\textbf{H1}). We model it reflectively because job satisfaction is a unidimensional attitude~\cite{Fernández-Macias2014}. Following previous research~\cite{sirgy2018work, haar2014outcomes}, we hypothesize that work-life balance is negatively associated with turnover intentions (\textbf{H2}). Since it is a unidimensional perception, we model it reflectively. A prior meta-analysis showed that job embeddedness predicts turnover intentions~\cite{jiang2012and}. We hypothesize that this also holds for software professionals (\textbf{H3}). We model job embeddedness as a formative construct because it includes two dimensions (on-the-job and off-the-job embeddedness), which are not interchangeable. For example, having strong social ties in one's community does not mean one fits in at work, and vice versa. 
Our first hypotheses for software professionals are therefore:

\renewcommand{\theenumi}{\textbf{H\arabic{enumi}}}
\begin{enumerate}
    \item Job satisfaction is negatively associated with turnover intentions. 
    \item Work-life balance is negatively associated with turnover intentions.
    \item Job embeddedness is negatively associated with turnover intentions.
\end{enumerate}
\renewcommand{\theenumi}{\arabic{enumi}}

Furthermore, prior studies~\cite{sirgy2018work, haar2014outcomes} found that when employers support employees in balancing their work roles with their family roles, employees are more satisfied with and embedded in their jobs, leading to hypotheses \textbf{H4 \& H5} concerning software professionals:

\renewcommand{\theenumi}{\textbf{H\arabic{enumi}}}
\begin{enumerate}
\setcounter{enumi}{3}
    \item Work-life balance is positively associated with job satisfaction.
    \item Work-life balance is positively associated with job embeddedness.
\end{enumerate}
\renewcommand{\theenumi}{\arabic{enumi}}    

Next, we looked for plausible antecedents of job satisfaction and job embeddedness. A seminal theory in psychology, self- determination theory (SDT)~\cite{deci2017self}, posits that individuals function optimally and are more satisfied when their needs for autonomy, competence, and relatedness are met. Thus, attitudes toward technical and sociotechnical infrastructure could act as proxy measures for the quality of work infrastructure that supports the previously mentioned psychological needs resulting in higher job satisfaction. This leads to hypotheses \textbf{(H6 \& H7)}. Prior work has investigated the role of tools and software development processes in affecting job satisfaction of software engineers~\cite{storey2019towards}. We used a recently published Software Infrastructure Attitude Scale (SIAS)~\cite{anon2025} with validated psychometric properties to measure software professionals' attitudes towards software tools and processes and modeled them as reflective constructs.

As above, prior work outside of software engineering has found that job satisfaction depends on the characteristics or perceived quality of the job~\cite{morris2010job}, which we model as a formative construct because it has five non-interchangeable dimensions (\textbf{H8}). Thus arriving at these hypotheses concerning software professionals:

\renewcommand{\theenumi}{\textbf{H\arabic{enumi}}}
\begin{enumerate}
\setcounter{enumi}{5}    
    \item Attitude towards sociotechnical software infrastructure is positively associated with job satisfaction.
    \item Attitude towards technical software infrastructure is positively associated with job satisfaction.
    \item Job quality is positively associated with job satisfaction.
\end{enumerate}
\renewcommand{\theenumi}{\arabic{enumi}}

Based on prior work~\cite{nguyen2017organizational}, we hypothesize that software professionals who perceive their workplaces as more just are more embedded in their jobs. Due to the different dimensions of organizational justice, we model it as a formative construct. This leads to the following hypothesis:

\renewcommand{\theenumi}{\textbf{H\arabic{enumi}}}
\begin{enumerate}
\setcounter{enumi}{8}    
    \item Organizational justice is positively associated with job embeddedness.
\end{enumerate}
\renewcommand{\theenumi}{\arabic{enumi}}

Meta-analyses~\cite{judge2002five, peltokorpi2023interaction} indicate that individual differences influence job satisfaction and job embeddedness, motivating hypotheses \textbf{H10–H15}. We focused on the traits extraversion, conscientiousness, and neuroticism, and excluded agreeableness and openness, based on Judge, Heller, and Mount~\cite{judge2000personality}, who report correlations of .20 or higher with job satisfaction for the former traits and .04 or lower for the latter. We model these personality traits as reflective constructs and hypothesize their associations with job satisfaction and job embeddedness.

\renewcommand{\theenumi}{\textbf{H\arabic{enumi}}}
\begin{enumerate}
    \setcounter{enumi}{9} 
    \item Extraversion is positively associated with job satisfaction.
    \item Extraversion is positively associated with job embeddedness.
    \item Conscientiousness is positively associated with job satisfaction.
    \item Conscientiousness is positively associated with job embeddedness.
    \item Neuroticism is negatively associated with job satisfaction.
    \item Neuroticism is negatively associated with job embeddedness.
\end{enumerate}
\renewcommand{\theenumi}{\arabic{enumi}}


\section{Methodology}\label{sec:methodology}

We conducted a cross-sectional survey study to test the hypotheses posed in the previous section. First, we define the population and inclusion criteria used in our study, followed by a description of the creation of the survey instrument, its piloting, and its deployment. At the end of the section we describe the data preprocessing, statistical procedures used in the analysis, the assessment of measurement validity for the reflective and formative constructs related to our hypotheses and the exploratory analysis conducted.

\subsection{Population and inclusion criteria}
The target population for this study is software professionals (e.g., programmers, product designers, product managers, quality assurance analysts, etc.) who are primarily employed in the creation and maintenance of software systems anywhere in the world. 



We excluded professionals who are primarily self-employed or freelancers (because they do not experience turnover in the same sense as regular employees), participants over the age of 65 (because retirement can confound voluntary job turnover), and participants under the age of 18 to avoid ethical concerns with research on minors.

\subsection{Survey instrument}

The full text of the questionnaire is available in the replication package (see \nameref{sec:data_availability}). The questionnaire begins with a consent form, followed by screening questions regarding age and employment.

Given the complexity of the model and natural limits of human attention, we opted for the shortest scales available that would still support assessing internal consistency. We measured turnover intentions using the Intent to Stay scale~\cite{mobley1978evaluation}, which has three items, including ``I often think of leaving the organization''. We measured overall job satisfaction with the Short Index of Job Satisfaction~\cite{judge2000personality}, which has 5 items. An example item from this scale is ''I feel real enjoyment at my work". Work-life balance was measured with the 3-item scale by Haar~\cite{haar2013testing}, with an example item ''I manage to balance the demands of my work and personal/family life 
well". Job embeddedness was measured using the Short Job Embeddedness scale~\cite{felps2009turnover}, which has 18 items. An example item is ``I really love the place where I live". Job quality was measured with the revised job diagnostic survey~\cite{idaszak1987revision}, which has 15 items for 5 dimensions. An example item is ``The job gives me considerable opportunity for independence and freedom in how I do the work". Dimensions of personality, that is extraversion, conscientiousness, and neuroticism, were measured with items from the Mini IPIP~\cite{donnellan2006mini}, which has 15 items. Organizational justice was measured with the short measure of organizational justice~\cite{elovainio2010developing}, which has 8 items in total. An example item is ''My supervisor has treated me with respect".

To improve data quality, all demographic questions were optional. We included an ``I don't know'' option for each scale item~\cite[see][]{oppenheim2000questionnaire, dolnicar2014including}. We also added an attention-check item to each of the two longest scales~\cite{kung2018attention}. The order of the items in each scale was randomized to reduce order effects~\cite{malhotra2008completion}. The survey was implemented using the software \textsf{Opinio}.\footnote{https://www.objectplanet.com/opinio/} The final instrument consisted of 77 items ,including the attention-check items. 

\subsection{Pilots}
When launching localized versions of the survey, the instrument was piloted with a handful of participants in each locale. We corrected some items translations for clarity and comprehension based on feedback. After initial data gathering, we found that filling the instrument takes between 10 and 15 minutes, largely depending on how much participants spend on optional demographic and open-ended questions.

\subsection{Deploying, sampling and localization}

\begin{table}[tb]
\centering
\caption{Self-reported demographic information of the respondents. All demographic questions but age were optional to answer.}
\begin{tabular}{llrr}
\toprule
& \textbf{Demographic Item} & \textbf{N} & \textbf{\%} \\
\midrule
\multirow{3}{*}{\textit{Gender}} 
    & Man & 157 & 70\% \\
    & Woman & 59 & 26\% \\
    & NA / Prefer not to say & 8 & 4\% \\
\midrule
\multirow{5}{*}{\begin{tabular}[c]{@{}l@{}}\textit{Country of}\\\textit{Residence (Top 5)}\end{tabular}} 
    & Brazil & 60 & 27\% \\
    & People's Republic of China & 31 & 14\% \\
    & Finland & 25 & 11\% \\
    & Thailand & 21 & 9\% \\
    & Saudi Arabia & 13 & 6\% \\
\midrule
\multirow{5}{*}{\begin{tabular}[c]{@{}l@{}}\textit{Survey Language}\\\textit{(Top 5)}\end{tabular}} 
    & Portuguese & 61 & 27\% \\
    & Chinese & 48 & 21\% \\
    & Thai & 27 & 12\% \\
    & Finnish & 23 & 10\% \\
    & English & 21 & 9\% \\
    & Other & 45 & 20\% \\
\midrule
\multirow{6}{*}{\begin{tabular}[c]{@{}l@{}}\textit{Age}\\\textit{(Range: 19--61)}\end{tabular}} 
    & Mean: 35.4, Median: 34 & & \\
    & 18--24 & 13 & 6\% \\
    & 25--30 & 66 & 29\% \\
    & 31--40 & 87 & 39\% \\
    & 41--50 & 39 & 17\% \\
    & 55 or more & 19 & 8\% \\
\midrule
\multirow{5}{*}{\begin{tabular}[c]{@{}l@{}}\textit{Years of}\\\textit{Experience}\\\textit{(Range: 0--40)}\end{tabular}} 
    & Mean: 10.4, Median: 8 & & \\
    & 0--4 years & 57 & 25\% \\
    & 5--9 years & 63 & 28\% \\
    & 10--19 years & 58 & 26\% \\
    & 20+ years & 33 & 15\% \\
\midrule
\multirow{4}{*}{\textit{Highest Education}} 
    & Doctorate & 15 & 7\% \\
    & Master's & 79 & 35\% \\
    & Bachelor's & 93 & 41\% \\
    & Other or NA & 37 & 16\% \\
\midrule
\multirow{7}{*}{\textit{Company Size}} 
    & 0--9 & 11 & 5\% \\
    & 10--99 & 46 & 20\% \\
    & 100--999 & 42 & 19\% \\
    & 1000--9999 & 69 & 31\% \\
    & 10,000--99,999 & 23 & 10\% \\
    & 100,000 or more & 14 & 6\% \\
    & NA & 19 & 8\% \\
\bottomrule
\end{tabular}
\label{tab:demo}
\end{table}

We report a post hoc power calculation using G*Power~\cite{kang2021sample} (version 3.1.9.7). Following Hair et al.~\cite{hair2021partial}, we based the calculation on the most complex regression in the model: predicting Job Satisfaction, which included six predictors and two control variables, as we removed conscientiousness from the model after for poorly loading items. Assuming an alpha level of 0.05 and a medium effect size ($f^2 = 0.15$), the analysis yielded an estimated power of 99\%, indicating a very high likelihood of detecting true effects of at least medium size. Even under a more conservative assumption of a small-to-medium effect ($f^2 = 0.10$), the model achieved a power of 93.7\%. These values exceed the commonly-accepted threshold of 0.80 for statistical power in behavioral and social science research~\cite{serdar2021sample}. 

After acquiring institutional ethics approval, the survey was localized to Bengali, Chinese, Farsi, Finnish, French, German, Japanese, Portuguese, Russian, Spanish, Swedish, Thai, and Urdu. Each translation was performed by a researcher with experience in conducting human factors research in software engineering. The study was advertised through our industry contacts, social media (e.g., LinkedIn), programming related events and communication channels such as Slack. Different incentive structures were applied depending on the locale where the study was advertised.

We received a total of 239 responses between December 12, 2024 and July 7, 2025. We removed 14 for failing one or both attention check items. Additionally, we removed one response due to variability (z = 3.64), which indicates non-engaged or random responding, bringing the total number of analyzable responses to 224. The self-reported demographic variables are shown in Table~\ref{tab:demo}. Both age and years of experience were collected as numeric responses but were later binned in the shared dataset to protect participant privacy. Other demographic questions were categorical. Note that none of the demographic questions other than age were mandatory to answer, meaning the number of responses from question to question changes. Respondents reported a wide range of software development experience (range of 0–40 years, mean = 10.4), with the majority holding a bachelor's (41\%) or master’s degree (36\%). The participants were employed in organizations of varying sizes, with nearly half of the respondents reporting to work in companies with less than 1000 employees. Respondents were from diverse geographic and linguistic backgrounds, with Brazil, China, and Finland being the most common countries of residence, and Portuguese, Chinese, and Thai among the most common languages the survey was taken in. 70\% of the respondents identified as man, 26\% as woman, 5 participants preferred not to say and 3 participants did not answer the question. The survey questionnaire also allowed the respondents to identify outside the gender binary, but no respondents selected it.

\subsection{Data preprocessing}

\begin{center}
\begin{table}
\centering
\caption{Number of Indicators (N), Scale Range (SR), Mean, Standard Deviation ($\sigma$), and Missingness (Miss \%) by Construct.}
\begin{tabular}{lccSSS}
\toprule
\textbf{Construct} & \textbf{N} & \textbf{SR} & \textbf{Mean} & \textbf{$\sigma$} & \textbf{Miss \%} \\
\midrule
Turnover Intentions    & 3  & 5 & 2.39 & 1.07 & 4.46 \\
Job Satisfaction      &  5 & 7 & 5.03 & 1.23 & 0.00 \\
Work-Life Balance      & 3  & 5 & 3.54 & 1.06 & 0.89 \\
Extraversion         & 4    & 5 & 2.82 & 0.96 & 1.34 \\
Conscientiousness    & 4     & 5 & 3.59 & 0.75 & 2.23 \\
Neuroticism         & 4   & 5 & 2.76 & 0.81 & 1.34 \\
Tech Attitude          & 6   & 5 & 3.75 & 0.79 & 2.23 \\
Socio Attitude         & 5  & 5 & 3.59 & 0.82 & 6.25 \\
\midrule
Autonomy              & 3    & 7 & 5.51 & 1.15 & 0.45 \\
Task Identity         & 3    & 7 & 5.22 & 1.24 & 3.24 \\
Task Significance          & 3    & 7 & 5.40 & 1.22 & 1.79 \\
Skill Variety         & 3    & 7 & 6.02 & 1.02 & 0.45\\
Feedback              & 3    & 7 & 5.31 & 1.16 & 2.23 \\
\midrule
Org. Links   & 3  & 5 & 4.04 & 0.73 & 1.79 \\
Org. Fit     & 3  & 5 & 3.73 & 0.85 & 3.12 \\
Org. Sacrifice& 3 & 5 & 3.60 & 0.87 & 2.23 \\
Community Links        & 3  & 5 & 2.84 & 1.07 & 1.34 \\
Community Fit         & 3   & 5 & 3.83 & 0.94 & 0.89 \\
Community Sacrifice   & 3   & 5 & 3.20 & 0.98 & 3.12 \\
\midrule
Procedural Justice   & 3    & 5 & 3.69 & 0.81 & 3.12 \\
Distributive Justice  & 2    & 5 & 3.28 & 1.15 & 4.02 \\
Interactional Justice & 3    & 5 & 4.28 & 0.78 & 6.25 \\
\midrule
\textbf{2\textsuperscript{nd} Order Constructs} & \textbf{N} & \textbf{SR} & \textbf{Mean} & \textbf{$\sigma$} & \textbf{Miss \%} \\
\midrule
Job Quality            & 15  & 7 & 5.49 & 0.87 & 6.70 \\
Job Embeddedness       & 18  & 5 & 3.52 & 0.58 & 10.71\\
Organizational Justice & 8  & 5 & 3.81 & 0.69 & 10.71 \\
\bottomrule
\end{tabular}
\label{tab:construct_descriptives}
\end{table}
\end{center}

Preprocessing the data involved reverse-scoring negatively worded items and coding ``I don’t know'' responses as missing (NA). Table~\ref{tab:construct_descriptives} reports descriptive statistics for all constructs. Missingness was generally low, with 53 indicators out of 75 having missingness less than 1\%, with a range of 0 to 6.25\%. For first-order constructs, the percentage of respondents with at least one missing item per construct was under 5.80\% and all but 4 constructs below 4\%. Among second-order constructs, the highest missing rate was 10.71\% for job embeddedness and Organizational Justice, resulting from having 18 and 8 indicators in total. For the rest of the analyses, we imputed missing values using predictive mean matching (PMM) using the R package mice~\cite{van2011mice} version 3.18.0. Regression based imputation techniques have been advocated for PLS-SEM imputation in general~\cite{amusa2024empirical}, as well as PMM in specific~\cite{reitz2025pursuing}. To include gender as a control variable in our model, we created a binary gender variable from participants’ self-reported gender. Respondents who identified as “Man” were coded as 0 and those who identified as “Woman” were coded as 1. Cases where gender was missing or marked as “Prefer not to say” (N=8) were set as missing (NA) in the binary variable. The resulting binary variable was then imputed as a categorical (factor) variable using the PMM imputation.

\subsection{Performing the analysis}

For our research, we used Partial Least Squares Structural Equation Modeling (PLS-SEM), because it deals with complex models with many indicators and latent constructs, and makes fewer assumptions about data distribution~\cite{rigdon2012rethinking, hair2019use}. Unlike covariance-based SEM, PLS-SEM offers greater flexibility in modeling hierarchical and formative constructs~\cite{wetzels2009using, becker2012hierarchical}. PLS-SEM also supports predictive modeling and is robust under smaller or heterogeneous samples~\cite{hair2019use}. For conducting the analysis, we followed the guidelines by Hair et al.~\cite{hair2021partial}

All analyses were conducted in R using the \texttt{cSEM} package (version 0.6.1)~\cite{rademaker2020cSEM}, including the assessment of reflective and formative constructs (Tables~\ref{tab:reliability} and~\ref{tab:formative_quality}), the bootstrapped structural model, $f^2$ effect sizes, and $R^2$ / adjusted $R^2$ values (Tables~\ref{tab:path_estimates} and~\ref{tab:r2_values}).

Model estimation followed a two-stage approach. In the first stage, constructs were specified as reflective or formative based on theoretical and empirical considerations. Higher-order formative constructs were modeled using the disjoint two-stage approach~\cite{sarstedt2019specify}: latent scores from first-order constructs were first estimated and then used as indicators in the second-order structural model.

Bootstrapping with 10{,}000 resamples was used in the second stage to assess the significance of path coefficients. Model fit was evaluated using standard indices, including SRMR, provided by the cSEM package. The bootstrapped model estimated formative second-order constructs using raw (i.e., non-disattenuated) first-stage scores, meaning measurement error was not corrected in the second-stage estimation~\cite{schuberth2020estimating}. This is also called the disjoint two-stage approach~\cite{sarstedt2019specify}.

\subsubsection{Reflective constructs}

All reflective constructs were evaluated for internal consistency and convergent validity using Cronbach’s alpha ($\alpha$), composite reliability ($\rho_C$), and average variance extracted (AVE), all from the bootstrapped model. All reliability metrics were calculated using the cSEM package~\cite{rademaker2020cSEM}. We also report Dijkstra-Henseler’s reliability ($\rho_A$) from the non-bootstrapped model, as it is not currently available in the cSEM package for the bootstrapped model. Following Hair et al.~\cite{hair2021partial}, we set acceptable thresholds to $0.70$ for $\alpha$, $\rho_C$, and $\rho_A$, and 0.50 for AVE. Constructs failing to meet these thresholds were considered for removal case-by-case basis, as explained below. Table~\ref{tab:reliability} shows reliability metrics for reflective constructs.

Main outcome variables in our model, turnover intentions, job satisfaction, and work-life balance showed good reliability, with $\alpha$ ranging from 0.77 to 0.89, $\rho_C$ between 0.87 and 0.94, AVEs between 0.69 and 0.83, and $\rho_A$ between 0.79 and 0.90. The hypothesized antecedents of job satisfaction, attitudes towards technical and sociotechnical infrastructure, showed excellent reliability as well.

Conscientiousness showed low reliability ($\alpha = 0.62$; $\rho_C = 0.33$; AVE = 0.17; $\rho_A = 0.49$). This could not be improved by dropping individual indicators, and therefore we excluded conscientiousness from our model. After removing one indicator for neuroticism, it showed borderline reliability ($\alpha = 0.69$; $\rho_A = 0.70$) and was kept in the model. We also removed two indicators for extraversion because of low loadings, and it demonstrated acceptable reliability, with $\alpha = 0.75$, $\rho_C = 0.85$, AVE = 0.75, and $\rho_A = 0.80$.

For job characteristic constructs, we removed indicators from task identity, autonomy and task significance due to low loadings. Afterwards all five core job characteristic constructs, that is autonomy, task identity, task significance, skill variety, and feedback, showed acceptable internal consistency and convergent validity, with AVEs between 0.62 and 0.78 and $\rho_C$ ranging from 0.83 to 0.88. Task identity showed borderline $\alpha$ value (0.68).

The organizational embeddedness dimensions showed more mixed results. Organizational fit performed well, with all reliability indices above the thresholds. Organizational sacrifice had a low $\alpha$ =0.62, but other reliability estimates over thresholds. Organizational links showed weak internal consistency ($\alpha = 0.47$) but acceptable composite reliability ($\rho_C = 0.78$), AVE = 0.63, and $\rho_A = 0.66$. Following Hair et al.~\cite{hair2021partial}, we retained organizational links based on composite reliability and the theoretical importance of capturing all dimensions of job embeddedness.

The community embeddedness dimensions also mostly met acceptable thresholds, with community fit meeting all thresholds and community sacrifice having borderline internal consistency ($\alpha = 0.69$). Community links was improved by removing one low loading indicator, though it still had very weak internal consistency ($\alpha = 0.47$, $\rho_A = 0.48$), but strong $\rho_C = 0.79$ and $AVE = 0.65$). As all metrics exceeded the minimum cutoff of 0.50 for AVE and 0.70 for $\rho_C$, warranting the retention of all constructs given their role in the nomological network of job embeddedness.

All constructs part of organizational justice met thresholds, with $\alpha$ ranging from 0.77 to 0.92, $\rho_C$ from 0.87 to 0.96, AVEs 0.69 to 0.93 and $\rho_A$ 0.78 to 0.91. Thus, the reliability metrics support the inclusion of nearly all reflective constructs in the model. Only conscientiousness was excluded based on low internal consistency and convergent validity.

\begin{table}[t]
\centering
\caption{Reliability Metrics for Reflective Constructs. }
\begin{tabular}{lcccc}
\toprule
\textbf{Construct} & \textbf{$\alpha$} & \textbf{$\rho_C$} & \textbf{AVE} & \textbf{$\rho_A$}\\
\midrule
Turnover Intentions  & 0.77 & 0.87& 0.69 & 0.79 \\
Job Satisfaction     & 0.88 & 0.92 & 0.74 & 0.89 \\
Work-Life Balance    & 0.89 & 0.94 & 0.83 & 0.90 \\
Tech Attitude        & 0.91 & 0.93 & 0.69 & 0.91 \\
Socio Attitude       & 0.89 & 0.91 & 0.69 & 0.89  \\
Neuroticism          & 0.69 & 0.83 & 0.61 & 0.70  \\
Extraversion & 0.75 & 0.85 & 0.75 & 0.80  \\
\midrule
Autonomy              &  0.73 & 0.88 & 0.78 & 0.73 \\
Task Identity          &  0.68 & 0.86 & 0.75 & 0.72 \\
Task Significance     & 0.73 & 0.87 & 0.77 & 0.77 \\
Feedback              &  0.70 &0.83& 0.62 & 0.70 \\
Skill Variety           & 0.79 & 0.88 & 0.70 & 0.80 \\
\midrule
Org. Fit      & 0.75 & 0.86 &0.67 &0.70 \\
Org. Links    & 0.49 & 0.78 &0.64 &0.66 \\
Org. Sacrifice& 0.62 & 0.79 &0.57 & 0.70 \\
Community Fit           & 0.75 &0.86 & 0.67 & 0.77 \\
Community Links         & 0.47 &0.79 & 0.65  &0.48 \\
Community Sacrifice     & 0.69 &0.83 & 0.62 & 0.70 \\
\midrule
Procedural Justice       & 0.77 &0.87 & 0.69 &0.78  \\
Distributive Justice     & 0.92 &0.96 & 0.93 & 0.91 \\
Interactional Justice    & 0.81 & 0.89 & 0.74 & 0.83 \\
\bottomrule
\end{tabular}
\label{tab:reliability}
\end{table}


\subsubsection{Formative Constructs}

We modeled \textit{Job Quality}, \textit{Job Embeddedness}, and \textit{Organizational Justice} as formative constructs based on conceptual justifications in Section~\ref{sec:background}. Internal consistency metrics such as $\alpha$, $\rho_C$, $\rho_A$, and AVE are not appropriate for formative constructs, because formative indicators are not expected to correlate highly~\cite{hair2021partial}. Thus we do not report these metrics for Job Quality, Job Embeddedness, and Organizational Justice.

We assessed formative construct quality using criteria proposed by Hair et al.~\cite{hair2021partial}, first, by estimating indicator weights and their significance through bootstrapping with 10{,}000 resamples. Next we assessed multicollinearity among indicators using variance inflation factors (VIF), with a recommended threshold of 3.3, which all indicators met.

\begin{table}[tbp]
\centering
\setlength{\tabcolsep}{3pt}
\caption{Formative constructs: Weights, significance, and VIF.}
\begin{tabular}{lrrrrr}
\toprule
\textbf{Indicator} & \textbf{Weight} & \textbf{$t$} & \textbf{95\% CI} & \textbf{$p$} & \textbf{VIF} \\
\midrule
\multicolumn{6}{l}{\textbf{Construct: Job Quality}} \\
 Autonomy     & 0.47  & 3.02  & [0.12; 0.72]  & 0.002    & 1.83 \\
 Task Identity     & 0.13  & 0.88  & [-0.17; 0.42]  & 0.38 & 1.46 \\
 Task Significance & 0.48  & 3.14 & [0.19; 0.78]  & 0.002    & 1.77 \\
 Skill Variety   & -0.30 & -1.50 & [-0.73; 0.06]  & 0.13 & 1.64 \\
 Feedback     & 0.36  & 2.56  & [0.07; 0.62]  & 0.01  & 1.77 \\
\midrule
\multicolumn{6}{l}{\textbf{Construct: Job Embeddedness}} \\
 Org. Links   & -0.04 & -0.44 & [-0.24; 0.13]  & 0.66      & 1.39 \\
 Org. Fit     & 0.63  & 6.98  & [0.44; 0.80]  & <0.001    & 2.37 \\
 Org. Sac.    & 0.50  & 5.32  & [0.31; 0.68]  & <0.001    & 2.11 \\
 Com. Links   & -0.13 & -1.90 & [-0.26; 0.01] & 0.06      & 1.27 \\
 Com. Fit     & 0.05  & 0.50  & [-0.18; 0.23]  & 0.62   & 1.58 \\
 Com. Sac.    & -0.12 & -1.51 & [-0.28; 0.04]  & 0.13  & 1.75 \\
\midrule
\multicolumn{6}{l}{\textbf{Construct: Organizational Justice}} \\
 Procedural     & 0.45  & 3.97  & [0.21; 0.66]  & 0.001 & 1.23 \\
 Distributive   & 0.46  & 4.53  & [0.25; 0.64]  & <0.001 & 1.28 \\
 Interactional  & 0.41  & 3.83  & [0.20; 0.62]  & 0.001 & 1.23 \\
\bottomrule
\end{tabular}
\label{tab:formative_quality}
\end{table}

\subsection{Exploratory indirect effects}

\textit{Mediation} occurs when the effect of an independent variable on a dependent variable is transmitted through an intervening variable, or \textit{mediator}~\cite{baron1986moderator}. In our proposed model, when we hypothesize a mediated effect (e.g. job satisfaction mediates the effect of job quality on intention to turnover), we did not include a direct effect (e.g. a path straight from job quality to turnover intentions with no mediator) because we found no theoretical basis on which to hypothesize these direct effects (work-life balance being the exception). However, determining whether to model a direct effect, and indirect effect, or both, is an important aspect of theory development; therefore, we report post-hoc sensitivity analysis of these non-hypothesized direct effects to provide a better starting point for future studies.

\section{Results}\label{sec:results}

\begin{figure*}[htp]
\centering
\includegraphics[width=0.9\textwidth]{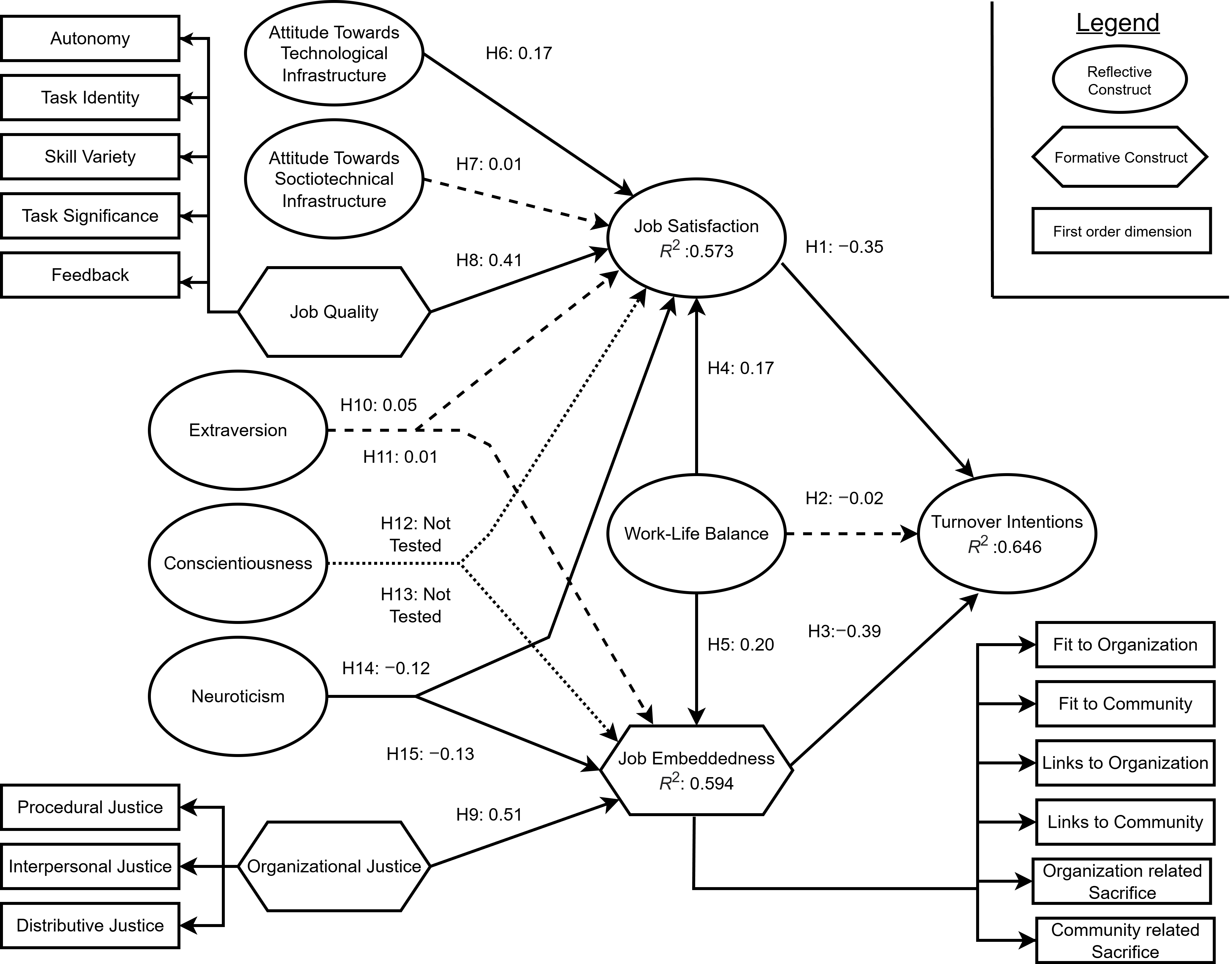}
\caption{Significant paths (p < .05) are indicated by full lines, non-significant paths with dashed lines and non-tested lines with dotted lines.} 
\label{fig:model}
\Description[Optional short description if the full description is longer than one sentence.]{The figure illustrates our structural equation model linking antecedents of turnover intentions to turnover intentions of software professionals. Job Satisfaction, Work–Life Balance, and Job embeddedness appear in the center of the model as mediators. Job satisfaction and work-life balance are modeled as a reflective constructs, job embeddedness as a formative construct. Arrows indicate hypothesized directional relationships, each labeled with its standardized path coefficient.

Job quality and organizational justice are formative constructs composed of several first-order dimensions (e.g., autonomy, task identity, feedback; procedural, interpersonal, and distributive justice). Attitudes toward technological and sociotechnical infrastructure, and the personality traits extraversion, conscientiousness, and neuroticism, are modeled as separate reflective constructs.

Job quality and organizational justice show the strongest positive paths to job satisfaction. Job satisfaction increases work–life balance and, together with work–life balance, contributes to job embeddedness. Neuroticism has small negative paths to work–life balance and job embeddedness. Extraversion and attitudes show weaker effects.

Turnover intentions is the main outcome. It is most strongly reduced by job embeddedness, followed by Job Satisfaction. The direct path from work–life balance to turnover intentions is non significant. A legend identifies construct types: circles for reflective, hexagons for formative, and rectangles for first-order dimensions.}
\end{figure*}

Table~\ref{tab:path_estimates} summarizes the estimated path coefficients, standard errors, t-values, and 95\% confidence intervals. Table~\ref{tab:r2_values} presents the $R^2$ and adjusted $R^2$ values for the outcome constructs in the non-bootstrapped and bootstrapped models. Figure~\ref{fig:model} depicts the hypothesized pathways, estimates, and coefficients of determination. Exploratory analysis on indirect effects is shown in Table~\ref{tab:indirect_effects}.

\subsection{Explained variance and model fit}

For transparency, we also report overall model fit using the fit indices. The standardized root mean square residual (SRMR) value for the model was 0.08, which is just at the commonly recommended threshold of 0.08~\cite{hair2021partial}. While this suggests moderate model fit, we note that the SRMR threshold is primarily derived from covariance-based SEM, and model fit in PLS-SEM should be interpreted with caution~\cite{henseler2013goodness, henseler2016using}. Additionally, the root mean square error of approximation (RMSEA) was 0.059, which is within the range commonly considered reasonable. RMSEA should be interpreted with similar caution as SRMR. Given the high explanatory power of the model (e.g., $R^2 = 0.656$ for turnover intentions) and the significance of key structural paths, we consider the model adequate for the purposes of this study.

Table~\ref{tab:r2_values} shows $R^2$ for both the bootstrapped and non-bootstrapped models. The non-bootstrapped model explained substantial variance in all outcome variables, with an $R^2$ of 0.646 for turnover intentions, 0.573 for job satisfaction, and 0.594 for job embeddedness. The $R^2$ of the bootstrapped model is lower ($R^2$ of 0.463 for turnover intentions, 0.457 for job satisfaction, and 0.453 for job embeddedness). These can be considered conservative and optimistic estimates. Table~\ref{tab:path_estimates} shows path estimates and $f^2$ effect sizes. We interpreted these values using Hair Jr.'s~\cite{hair2019use} thresholds: values higher than 0.02, 0.15, and 0.35 correspond to small, medium and large effects. $f^2$ represents the unique contribution of each predictor to the variance explained in an endogenous construct. As such, predictors in models with multiple predictors may exhibit lower $f^2$ values due to shared explained variance among predictors.

\begin{table}[tbp]
\centering
\setlength{\tabcolsep}{3pt}
\caption{Bootstrapped path coefficients, and confidence intervals (CI). Effect sizes ($f^2$) from the non-bootstrapped model.}
\label{tab:path_estimates}
\begin{tabular}{lrrrrrr}
\toprule
\textbf{Path} & \textbf{Est.} & \textbf{SE} & \textbf{$t$} & \textbf{95\% CI}& \textbf{$p$} & \textbf{$f^2$} \\
\midrule
\multicolumn{6}{l}{\textbf{Outcome: Turnover Intentions}} \\
Job Satisfaction        & -0.35 & 0.08 & -4.30 & [-0.48; -0.17] & <0.001 &0.11 \\
Work-Life Balance       & -0.02 & 0.07 & -0.23 & [-0.14; 0.12] & 0.81 &0.00 \\
Job Embeddedness        & -0.39 & 0.09 & -4.28 & [-0.59; -0.25] & <0.001&0.29 \\

\midrule
\multicolumn{6}{l}{\textbf{Outcome: Job Satisfaction}} \\
Tech Attitude           &  0.17 & 0.07 &  2.59 & [0.03; 0.30]   & 0.01 &0.04 \\
Socio Attitude          &  0.01 & 0.07 & 0.19 & [-0.11; 0.14]  & 0.85 &0.00 \\
Job Quality             &  0.41 & 0.06 &  6.60 & [0.30; 0.55]   & <0.001&0.43 \\
Work-Life Balance       &  0.17 & 0.06 &  2.78 & [0.05; 0.29]   & 0.005 &0.03 \\
Neuroticism             & -0.12 & 0.06 & -2.02 & [-0.24; -0.00] & 0.04 &0.04 \\
Extraversion            &  0.05 & 0.06 &  0.77 & [-0.07; 0.16]  & 0.44 &0.01 \\
Age                     &  0.12 & 0.05 &  2.41 & [0.01; 0.21]   & 0.02 &0.03 \\
Gender                  &  -0.02 & 0.05 &  -0.42 & [-0.13; 0.08]  & 0.67 &0.00 \\

\midrule
\multicolumn{6}{l}{\textbf{Outcome: Job Embeddedness}} \\
Work-Life Balance       &  0.20 & 0.07 &  2.77 & [0.04; 0.33]   & 0.006 &0.04 \\
Org. Justice  &  0.51 & 0.06 &  8.55 & [0.40; 0.63]   & <0.001 &0.59 \\
Neuroticism             & -0.13 & 0.07 & -1.93 & [-0.25; 0.01]  & 0.05 &0.05 \\
Extraversion            &  -0.04 & 0.06 &  0.58 & [-0.16; 0.09]  & 0.56 &0.01 \\
Age                     &  0.00 & 0.07 &  -0.03 & [-0.14; 0.13]  & 0.98 &0.00 \\
Gender                  &  0.03 & 0.05 &  0.68 & [-0.06; 0.12]  & 0.49 &0.00 \\
\bottomrule
\end{tabular}
\end{table}

\begin{table}[t]
\centering
\caption{Bootstrapped R-squared for Endogenous Constructs.}
\label{tab:r2_values}
\begin{tabular}{lrrrr}
\toprule
\textbf{Construct} & \textbf{$R^2$} & \textbf{Adj. $R^2$} & \textbf{B. $R^2$} & \textbf{B. Adj. $R^2$}\\
\midrule
Turnover Intentions & 0.646 & 0.642 & 0.463  & 0.456\\
Job Satisfaction & 0.573 & 0.556  &  0.457 &  0.437\\
Job Embeddedness & 0.594 &  0.586 & 0.453 & 0.438 \\
\bottomrule
\end{tabular}
\end{table}

\begin{table}[ht]
\centering
\caption{Exploratory indirect effects (via mediators) on Turnover Intentions.}
\label{tab:indirect_effects}
\begin{tabular}{lrrrrr}
\toprule
  \textbf{Path} & \textbf{Est} & \textbf{Std. E} & \textbf{$t$} & \textbf{95\% CI} &  \textbf{$p$} \\
\midrule
Org. Justice & \textbf{-0.20} & \textbf{0.06} & \textbf{-3.21} & \textbf{[-0.35;-0.11]} &\textbf{0.001} \\
Job Quality & \textbf{-0.14} & \textbf{0.03} & \textbf{-4.10} & \textbf{[-0.21;-0.07]}&\textbf{< 0.001} \\
Work-Life B. & \textbf{-0.14} & \textbf{0.04} & \textbf{-3.41} & \textbf{[-0.22;-0.06]} &\textbf{< 0.001} \\
Tech Attitude  & \textbf{-0.06} & \textbf{0.03}& \textbf{-2.27}& \textbf{[-0.11; -0.01]} &\textbf{0.02} \\
Neuroticism& \textbf{0.09} & \textbf{0.04} & \textbf{2.33} & \textbf{[0.01; 0.17]} & \textbf{0.02}\\
Socio Attitude&  -0.00 & 0.02&  -0.20 &  [-0.05; 0.03] &0.84 \\
Extraversion & -0.01 &  0.04 &  -0.05 & [-0.08; 0.07] &0.97 \\
\bottomrule
\end{tabular}
\end{table}

\subsection{Turnover intentions}

Three hypotheses (H1–H3) examined predictors of turnover intentions of software professionals. Job satisfaction was negatively associated with turnover intentions ($\beta = -0.33, t= -4.3$, 95\% CI [$-0.49, -0.16$]), with a small to medium effect size ($f^2 = 0.11$), supporting hypothesis \textbf{H1}. Job embeddedness showed a significant negative relationship with turnover intentions ($\beta= -0.39, t= -4.28$, 95\% CI [$-0.59, -0.25$]), with a medium to large effect size ($f^2 = 0.29$), supporting \textbf{H3}. Work-life balance was not directly associated with turnover intentions ($\beta= -0.02, t= -0.23$, 95\% CI [$-0.28, -0.01$]), thus \textbf{H2} was not supported.

\subsection{Job satisfaction}
Altogether seven hypotheses (H4, H6-H9, H12, and H13) investigated the predictors of job satisfaction. Work-life balance ($\beta= 0.17, t= 2.78$, 95\% CI [$0.05, 0.29$]) and job quality ($\beta= 0.41, t= 6.6$, 95\% CI [0.30, 0.55]) had statistically significant positive effects, with unique effects of small and medium ($f^2 = 0.04$ and $0.43$), supporting \textbf{H4} and \textbf{H8}, respectively. Attitude toward technical infrastructure ($\beta= 0.17,t= 2.59$, 95\% CI [$0.03, 0.30$]) also had a significant relationship with job satisfaction with a small effect size supporting \textbf{H6}. Sociotechnical infrastructure was not a significant predictor, and thus \textbf{H7} was not supported.

Neuroticism was negatively associated with job satisfaction ($\beta=-0.12,t=2.02$, 95\% CI [$-0.24, -0.00$]), with a small unique effect ($f^2 = 0.04$) supporting hypothesis \textbf{H14}. The relationship between extraversion and job satisfaction was not significant, thus not supporting \textbf{H10}. Hypothesis \textbf{H12} was not tested because of the poor reliability of the conscientiousness construct.

\subsection{Job embeddedness}
Five hypotheses (H5, H9, H11, H13 and H15) examine predictors of job embeddedness. Organizational justice was very strongly associated with job embeddedness ($\beta=0.51, t=8.55$, 95\% CI [$0.40, 0.63$]), with a large unique effect ($f^2 = 0.43$), supporting \textbf{H9}. Work-life balance was also associated with job embeddedness ($\beta =0.20, t=2.77$, 95\% CI [$0.04, 0.33$]) supporting \textbf{H5}. Neuroticism was associated with job embeddedness ($\beta =-0.13, t=-1.93$, 95\% CI [$-0.25, 0.01$]), with a small unique effect ($f^2 = 0.05$), supporting \textbf{H15}. Extraversion did not have a relationship with job embeddedness. Conscientiousness was not included in the model and thus \textbf{H13} was not tested.

\subsection{Indirect effects and control variables}
In an exploratory analysis of indirect effects, we examined how several predictors influence turnover intentions through mediating variables. The results, shown in Table~\ref{tab:indirect_effects}, indicate that four constructs had significant indirect effects on turnover intentions. Attitude toward technical infrastructure showed a significant negative indirect effect on turnover intentions ($\beta = -0.06, p = 0.02$), suggesting that favorable perceptions of technical infrastructure reduce turnover intentions via the intervening variable of job satisfaction. Job quality showed a significant negative indirect effect on turnover intentions ($\beta = -0.14, p = < 0.0001$), suggesting that individuals perceiving higher job quality are more likely to be satisfied with their job, thereby reducing their intent to leave. Neuroticism had a positive indirect effect ($\beta = 0.09, p = 0.02$), indicating that higher levels of neuroticism may increase turnover intentions indirectly. Extraversion did not show a significant indirect effect. Both work-life balance and organizational justice exhibited strong negative indirect effects on turnover intentions ($\beta = -0.14$ and $\beta = -0.20$, $p < 0.001$ and $p = 0.001$, respectively), suggesting that higher work-life balance and organizational justice reduce turnover intentions through increased job satisfaction and job embeddedness.

Control variable of age had a small effect on job satisfaction ($\beta = 0.12$, $p = 0.03$), meaning older people had higher job satisfaction. Gender did not have an effect on job satisfaction. Neither of the control variables had an effect on job embeddedness. 

\subsection{Formative construct evaluation}

Table~\ref{tab:formative_quality} reports the outer weights, confidence intervals, and VIF values for all formative constructs. For job quality, three indicators were statistically significant, with {autonomy ($\beta = 0.47$), feedback ($\beta = 0.36$), and task significance ($\beta = 0.48$) contributing most strongly to the construct. Of note is also the negative estimate for skill variety, which is, however not statistically significant. VIF values ranged from 1.46 to 1.83, well below the threshold of 3.3, indicating acceptable levels of multicollinearity.

For job embeddedness, the two organizational indicators (fit and sacrifice) showed significant contributions ($\beta = 0.63, 0.50$). However, community-related indicators were non-significant. This pattern may reflect domain-specific differences in what embeds software professionals to their jobs, but further research on this is needed. VIF values were acceptable with range of 1.27 to 2.37.

All three components of organizational justice, that is procedural, distributive, and interactional justice, were statistically significant, with VIF values well below the threshold (1.23–1.28), supporting the stability of the construct.
\section{Discussion}\label{sec:discussion}

Our proposed model of turnover intentions improves upon prior models. Our bootstrapped and non-bootstrapped models explain 46\% and 65\% of the variance in turnover intentions respectively, compared to 19--33\% for intent to stay at a job and turnover intentions in prior work~\cite{trinkenreich2024predicting, sharma2020exploring}. This represents absolute improvements of 32 to 46 percentage points for the non-bootstrapped model and 13 to 27 percentage points for the bootstrapped model, or relative improvements of 247\% and 97\% for the non-bootstrapped model and 142\% and 39\% for the bootstrapped model. While the prior work drew primarily on organizational theories such as the job demands–resources model and onboarding success, we drew on behavioral theories from social psychology, such as the theory of reasoned action and the theory of planned behavior. Our results thus suggest that these theories, which emphasize the role of individual attitudes and subjective norms in shaping intentions and behavior, explain more of the turnover intentions of software professionals.

Our findings are broadly in line with prior work conducted in organizational psychology with non-software professional samples. Interestingly, however, the coefficient for job embeddedness ($\beta=-0.39$) is higher than that for job satisfaction ($\beta=-0.35$), meaning embeddedness explains at least as much as job satisfaction in intent to change jobs. 

Similarly, job satisfaction is positively associated with work-life balance and job quality, and negatively with neuroticism. Job embeddedness is positively associated with perceived organizational justice and work-life balance, and negatively with neuroticism. Most notably, work-life balance does not directly affect turnover intentions, instead its influence is mediated through increased job satisfaction and job embeddedness.

Interestingly, however, attitudes toward technical infrastructure (e.g., software tools) are associated with job satisfaction, while attitude towards sociotechnical infrastructure (e.g., practices that organize work) is not. This is partly in line with the common wisdom of industry and practitioners, as reflected in the StackOverflow developer survey\footnote{https://survey.stackoverflow.co/2024/}, which receives tens of thousands of responses and asks for things such as which technologies are most admired and which of them do the developers want to work with. Our results are partly line with previous SE research linking tools and processes job satisfaction indirectly through perceived productivity~\cite{storey2019towards}.

\subsection{Implications}

Our results suggest that software companies aiming to reduce voluntary turnover should prioritize enhancing job satisfaction and job embeddedness. Together, these constructs explain 45–65\% of the variance in turnover intentions in our model. 

Job satisfaction was positively influenced by job quality, work-life balance and attitudes toward technical infrastructure. Among the dimensions of job quality, autonomy, task significance, and feedback from the work itself had the most impact. Interestingly, skill variety had a negative non-significant weight on job quality, implying that software professionals prefer to focus more on specific tasks. This might be a difference between software development, where too much skill variety might increase cognitive load and reduce focus, and industrial work, where skill variety brings welcome breaks from monotone tasks. This result warrants further research.

For the job embeddedness construct, the dimensions of organizational fit and sacrifice mattered the most, while none of the dimensions related to community were important for software professionals. Thus, it seems that software professionals are mainly embedded through organizational fit and perceived sacrifice of leaving a job, while a geographic location or community does not seem to matter. On the other hand, job embeddedness was strongly predicted by organizational justice, with all procedural, distributive, and interactional dimensions contributing meaningfully. In practice, this means that employees are more embedded to their jobs, if they perceive that the decisions made at work leading to outcomes are made fairly, when they feel that they are treated with respect, and when they feel that decision regarding outcomes and resources (e.g., pay and perks) are distributed fairly.

Thus, in conclusion, software companies aiming to increase employee retention should offer jobs that are experienced as meaningful, which offer autonomous work roles, and where the employee perceives how well they are performing in the job with feedback from the job itself. These jobs should be in organizations that promote work-life balance, where employees feel that decisions about work outcomes are made fairly, where their managers treat them with respect and where they feel that the compensation they receive is fair. Employees having a positive perception about the technical tools at their work can also experience lower turnover intentions. 

These findings have implications for research and education. For researchers, they highlight the value of studying psychological constructs (e.g., job embeddedness, justice perceptions) in the software industry. Future work could examine how these dynamics vary across organizational contexts and roles, and examine why different dimensions of these constructs are important to software professionals. 


\subsection{Limitations}

\textbf{Construct validity.} Reliability metrics, including $\alpha$, $\rho_C$, and $\rho_A$, exceeded recommended thresholds for the majority of constructs ($\alpha$, $\rho_C$, and $\rho_A > 0.70$; AVE > 0.50)~\cite{hair2021partial, henseler2016using}. Exceptions include conscientiousness, which was removed from the model, and neuroticism, organizational links, and community links, which, while below the threshold of 0.70 for $\alpha$, were kept in the model because $\rho_C$ and AVE values were within acceptable bounds and their theoretical importance to the formative construct of job embeddedness. The steps we took to improve construct validity included using broadly accepted scales and removing poorly loading items. Localization of the scales may have created construct validity issues for the constructs with low $\alpha$ values.

\textbf{Internal validity.} Due to the cross-sectional nature of our design, we cannot make causal claims. While structural equation modeling allows for testing directional relationships, causal inference requires longitudinal or experimental designs~\cite{shadish2002experimental}. Moreover, all data were collected via self-report in a single session, raising the possibility of common method variance (CMV). We randomized item order within scales, but some CMV may remain~\cite{podsakoff2003common}. We also chose to forego recruitment from crowdsourcing platforms due to reported data quality issues in software engineering research~\cite{reid2022software, russo2022recruiting}. We have discussed other challenges arising from data gathering in a previous publication~\cite{kuutila2025methodological}.

\textbf{External validity.} The generalizability of our findings is limited by our sample, as it is in all software engineering research, where no true random samples exist~\cite{baltes2022sampling}. While our participants were software professionals, our sample can over-represent geographic regions and company sizes. However, by localizing our survey into multiple languages, we have tried to make our results more generalizable by including data from diverse sources.

\textbf{Statistical conclusion validity.} While our model explained substantial variance in key outcomes, some paths were non-significant, and our sample size is not sufficient to detect small effects. Nevertheless, the use of PLS-SEM enhances statistical conclusion validity by allowing simultaneous estimation of complex models that include both measurement and structural components. This approach incorporates measurement error, tests all hypothesized relationships in a unified model, and is less sensitive to violations of multivariate normality and to smaller sample sizes~\cite{hair2021partial}. Additionally, model fit indices such as SRMR and RMSEA were reported for transparency. However, these indices are derived from covariance-based SEM traditions, and their applicability in PLS-SEM remains debated~\cite{hair2021partial, henseler2016using}.

\subsection{Future work}\label{ssec:futurework}
In the future, several aspects of our model can be further refined. First, while turnover intentions have been shown to correlate well with actual voluntary turnover, actual measured turnover could be added to our model as a direct measure. Secondly, behaviors such as searching for jobs, have been shown to correlate strongly with turnover intentions and actual turnover. This would increase the explanatory power of the model, while having more limited real world implications for software companies. Third, group differences, such as gender, age, and experience can be analyzed to provide further insights for the software industry. Fourth, longitudinal studies allow assessment of the effects of change in constructs (e.g., job satisfaction), rather than the absolute level, in influencing turnover intentions.

\section{Conclusions}\label{sec:conclusions}
To briefly answer our research question, our findings show that turnover intentions among software professionals are primarily explained by job satisfaction and job embeddedness, which in turn are primarily shaped by job quality, organizational justice, work-life balance and attitudes toward technical infrastructure. Thus, this study offers a theoretically and empirically grounded examination of turnover intentions among software professionals using partial least squares structural equation modeling (PLS-SEM). Our model explained a substantial proportion of variance in key outcomes, particularly turnover intentions ($R^2 = 0.646$), and revealed several significant pathways.

Job satisfaction and job embeddedness emerged as strong negative predictors of turnover intentions, reinforcing findings from organizational psychology within the software engineering context. Job satisfaction was positively predicted by job quality, attitude toward technical infrastructure and work-life balance, and negatively by neuroticism. Job embeddedness was strongly predicted by organizational justice, with community-related indicators showing little influence. This pattern suggests that software professionals seem to be more anchored by workplace factors than by local or community-based ties. This could be a domain specific dynamic and requires further investigation.

Exploratory analysis of indirect effects further showed that job quality, work-life balance, and organizational justice, among others, influenced turnover intentions via job satisfaction and job embeddedness. These mediation paths underscore the value of treating employees well and offering them jobs with autonomy, task significance, and feedback from the job itself to lessen turnover intentions.

By integrating established psychological theories with constructs relevant to software engineering, this study provides a nuanced, evidence-based perspective on voluntary turnover. Our findings are discussed with actionable insights for organizations aiming to improve retention and to inform future research on the human factors shaping software professionals' careers.
\section*{Data Availability} \label{sec:data_availability}
A replication package comprising (1) a dataset including data from participants who consented to their data being shared, (2) our analysis scripts, and (3) our survey instrument is available at Zenodo~\cite{anonymous2025}. 

\begin{acks}
This project was supported by National Sciences and Engineering Research Council of Canada Discovery Grant RGPIN-2020-05001, Discovery Accelerator Supplement RGPAS-2020-00081, and the Killam Foundation.
\end{acks}

\bibliographystyle{ACM-Reference-Format}
\bibliography{sample-base}

\end{document}